\documentclass[aps,prd,twocolumn,nofootinbib,superscriptaddress]{revtex4-1}
\usepackage[utf8]{inputenc}

\usepackage{color}
\usepackage{graphicx}
\usepackage{amsmath}
\usepackage{amssymb}
\usepackage{bm}
\usepackage{acronym}
\usepackage{ifthen}
\usepackage{blindtext}
\usepackage[normalem]{ulem}
\usepackage{hyperref}
\usepackage{etoolbox}
\usepackage{fancyhdr}
\usepackage{xspace}
\usepackage{textcomp}
\usepackage{multirow}
\usepackage[caption=false]{subfig}
\usepackage{lineno}
\usepackage{tabularx}

\def\gwh{gravitational-wave\xspace}
\def\gws{gravitational waves\xspace}

\newacro{S/N}{signal-to-noise ratio}

\newacro{PN}{post-Newtonian}

\newacro{O3}{the third observing run}

\newacro{O2}{the second observing run}

\newacro{CW}{\emph{Continuous Wave}}

\newcommand{\TFFT}{T_\text{FFT}}
\newcommand{\tfftmax}{T_\text{FFT,max}}

\newcommand{\bea}{\begin{eqnarray}}
\newcommand{\eea}{\end{eqnarray}}
\newcommand{\be}{\begin{equation}}
\newcommand{\ee}{\end{equation}}

\newtoggle{fullauthorlist}
\toggletrue{fullauthorlist}
\newtoggle{endauthorlist}
\toggletrue{endauthorlist}

\begin{document}
\title{Constraints on dark photon dark matter using data from LIGO's and Virgo's third observing run}
\date{\today}

\iftoggle{endauthorlist}{
  %
  %
  \let\mymaketitle\maketitle
  \let\myauthor\author
  \let\myaffiliation\affiliation
  \author{The LIGO Scientific Collaboration}
  \author{The Virgo Collaboration}
  \author{The KAGRA Collaboration}
  \email{Full author list given at the end of the article.}
\noaffiliation
}{
\iftoggle{fullauthorlist}{
  \input{KAGRA-LSC-Virgo-Authors-Aug-2020-aas}
}
{
  \author{The LIGO Scientific Collaboration}
  \affiliation{LSC}
  \author{The Virgo Collaboration}
  \affiliation{Virgo}
  \author{The KAGRA Collaboration}
  \affiliation{KAGRA}
  
}
}


\begin{abstract}
We present a search for dark photon dark matter that could couple to gravitational-wave
interferometers using data from Advanced LIGO and Virgo's third observing run. To perform this analysis,
we use two methods, one based on cross-correlation of the strain channels in the two nearly aligned LIGO detectors, and one that looks for excess power in the strain channels of the LIGO and Virgo detectors. The excess power method optimizes the Fourier Transform coherence time as a function of frequency, to account for the expected signal width due to Doppler modulations. We do not find any evidence of dark photon dark matter with a mass between
$m_{\rm A} \sim 10^{-14}-10^{-11}$ eV/$c^2$, which corresponds to frequencies between 10-2000 Hz, and therefore provide upper limits on the square of the minimum coupling of dark photons to baryons, i.e. $U(1)_{\rm B}$ dark matter. For the cross-correlation method, {the best median constraint on the squared coupling is {$\sim2.65\times10^{-46}$} at {$m_{\rm A}\sim4.31\times10^{-13}$} eV/$c^2$}; for the other analysis,  the best constraint is {$\sim 2.4\times 10^{-47}$} at {$m_{\rm A}\sim 5.7\times 10^{-13}$} eV/$c^2$. These limits improve upon those obtained in direct dark matter detection experiments by a factor of {$\sim100$} for {$m_{\rm A}\sim [2-4]\times 10^{-13}$} eV/$c^2$, {and are, in absolute terms, the most stringent constraint so far in a large mass range $m_A\sim$ $2\times 10^{-13}-8\times 10^{-12}$ eV/$c^2$.}
\end{abstract}
        
        \maketitle

\section{Introduction}\label{sec:intro}

Dark matter has been known to exist for decades \cite{Bertone:2016nfn}, yet its physical nature has remained elusive. Depending on the theory, dark matter could consist of particles with masses as low as $10^{-22}$ eV$/c^2$ \cite{Hui:2016ltb}, or as high as (sub-) solar-mass primordial black holes \cite{Bird:2016dcv,clesse2018seven,Miller:2020kmv,Miller:2021knj}. Furthermore, dark matter clouds could form around black holes that deplete over time and emit gravitational waves \cite{Brito:2015oca,isi2019directed}. Here, we focus on a subset of the ``ultralight'' dark matter regime, i.e. masses of $\mathcal{O}(10^{-14}-10^{-11})$ eV$/c^2$ \cite{bertone2019gravitational},
in which a variety of dark matter candidates may interact with \gwh interferometers. Scalar, dilaton dark matter could change the mass of the electron and other physical constants, causing oscillations in the Bohr radius of atoms in various components of the interferometer \cite{grote2019novel}; axions \cite{Peccei:1977hh} could alter the phase velocities of circularly polarized photons in the laser beams traveling down each arm of the detector \cite{nagano2019axion}; dark photons could couple to baryons in the mirrors, causing an oscillatory force on the detector \cite{PierceRilesZhao2018}; tensor bosons could also interact with the interferometer in an analogous way as \gws \cite{Armaleo:2020efr}. {Here, we focus on dark photon dark matter whose relic abundance could be induced by } the misalignment mechanism \cite{nelson2011dark,arias2012wispy,graham2016vector}, the tachyonic instability of a scalar field \cite{agrawal2020relic,pierce2019dark,bastero2019vector,dror2019parametric}, or cosmic string network decays \cite{long2019dark}. {Cosmic strings, in particular, also offer a promising way to probe physics beyond the standard model with \gwh detectors at energies much larger than those attainable by particle accelerators \cite{LIGOScientific:2021nrg}, which complements the kind of direct dark matter search we perform here. } {Independently of the formation mechanism, analyses of \gwh data could make a statement on the existence of dark photons.}

A search for dark photons using data from Advanced LIGO/Virgo's first observing run \cite{PierceRilesZhao2018,guo2019searching} has already been performed, resulting in competitive constraints on the coupling of dark photons to baryons. Furthermore, scalar, dilaton dark matter was searched for recently using data from GEO600 \cite{Dooley:2015fpa}, and upper limits were placed on the degree to which scalar dark matter could have altered the electron mass or fine-structure constant \cite{Vermeulen:2021epa}.

Other experiments that have probed the ultralight dark matter regime include the Eötvös experiment, which aims to find a violation to the equivalence principle of General Relativity resulting from a new force acting on test masses in a dark matter field, by looking for a difference in the horizontal accelerations of two different materials using a continuously rotating torsion balance \cite{Su:1994gu,Schlamminger:2007ht}; the MICROSCOPE satellite \cite{Touboul:2012ui}, which measures the accelerations of two freely-floating objects in space made of different materials to look for a violation of the equivalence principle and hence a new force \cite{Berge:2017ovy}; the Axion Dark Matter Experiment (ADMX), which searches for $\mathcal{O}(\mu$eV/$c^2)$ dark matter by trying to induce an axion-to-photon conversion in the presence of a strong magnetic field in a resonant cavity \cite{Du:2018uak}; 
and the Any Light Particle Search (ALPS), which looks for particles with masses less than $\mathcal{O}$(meV/$c^2$) (that could compose dark matter) by subjecting photons to strong magnetic fields in two cavities, separated by an opaque barrier, to cause a transition to an axion and then back to a photon \cite{Hartman:2020bgj}. 
Ultralight dark matter has also been constrained by observing gravitational waves from depleting boson clouds around black holes \cite{palomba2019direct,Sun:2019mqb,isi2019directed,d2018semicoherent,Ng:2020ruv,Zhu:2020tht,Tsukada:2018mbp}, and by analyzing binary mergers, e.g. GW190521, which is consistent with the merger of complex vector boson stars \cite{CalderonBustillo:2020srq}.

{Compared to the analysis on data from LIGO/Virgo's first observing run \cite{guo2019searching}, } we use two methods, one based on cross-correlation \cite{PierceRilesZhao2018}, and another that judiciously varies the Fourier Transform coherence time \cite{piccinibsd,Miller:2020vsl}, to search for dark photons in Advanced LIGO and Virgo data from the third observing run (O3). {Additionally, we include the signal induced by the common motion of the mirrors \cite{Morisaki:2020gui}- see section \ref{sec:model}. } Although we do not find any evidence for a dark photon signal, we place stringent upper limits on the degree to which dark photons could have coupled to the baryons in the interferometer.

\section{Dark matter interaction model} \label{sec:model}

Ultralight dark photon dark matter is expected to cause time-dependent oscillations in the mirrors of the LIGO/Virgo interferometers, which would lead to a differential strain on the detector. We formulate dark photons in an analogous way to ordinary photons: as having a vector potential with an associated dark electric field that causes a quasi-sinusoidal force on the mirrors in the interferometers. The Lagrangian $\mathcal{L}$ that characterizes the dark photon coupling to a number current density $J^\mu$ of baryons or baryons minus leptons is:

\begin{equation}
    \mathcal{L} = -\frac{1}{4\mu_0} F^{\mu \nu} F_{\mu \nu} + \frac{1}{2\mu_0} \left(\frac{m_A c }{\hbar}\right)^2 A^\mu A_\mu - \epsilon e J^\mu A_\mu, \label{lagrangian} \\
\end{equation}
where $F_{\mu \nu} = \partial_\mu A_\nu - \partial_\nu A_\mu$ is the electromagnetic field tensor, $\hbar$ is the reduced Planck's constant, $c$ is the speed of light, $\mu_0$ is the magnetic permeability in vacuum, $m_A$ is the dark photon mass, $A_\mu$ is the four-vector potential of the dark photon, $e$ is the electric charge, and $\epsilon$ is the strength of the particle/dark photon coupling normalized by the electromagnetic coupling constant. \footnote{We note that the dark photon in our scenario is a different from the one which couples to the standard model via kinetic mixing.}

If the analysis observation time exceeds the signal coherence time, given by equation 3 \cite{Miller:2020vsl}, we can write the acceleration of the identical LIGO/Virgo mirrors in the dark photon field as \cite{guo2019searching}:

\begin{equation}
 \vec{a}(t,\vec{x})\simeq\epsilon e \frac{q}{M}\omega \vec{A}\cos(\omega t -\vec{k} \cdot\vec{x}+ \phi)
\label{eqnacce}
 \end{equation}
where $\omega$, $\vec{k}$, and $\vec{A}$ are the angular frequency, propagation vector, and polarization vector of the dark photon field, $\vec{x}$ is the position of a mirror, $\phi$ is a random phase, and $q$ and $M$ are the charge and the mass of the mirror, respectively. If the dark photon couples to the baryon number, $q$ is the number of protons and neutrons in each mirror. If it couples to the difference between the baryon and lepton numbers, $q$ is the number of neutrons in each mirror. For a fused Silica mirror, $q/M=5.61\times 10^{26}$ charges/kg for baryon coupling and $q/M=2.80\times 10^{26}$ charges/kg for baryon-lepton coupling. 
Practically, we cannot distinguish between the two types of coupling, though the baryon-lepton coupling would lead to half the acceleration relative to that of the baryon coupling.
 
Because we observe for almost one year, significantly longer than the assumed dark photon coherence time, and the dark photons travel with non-relativistic velocities, we model the signal as a superposition of many plane waves, each with a velocity drawn from a Maxwell-Boltzmann distribution  \cite{carney2019ultralight}. 
The superposition of dark photon plane waves with different velocities leads to a frequency variation of the signal \cite{Miller:2020vsl,PierceRilesZhao2018}:

\be
\Delta f=\frac{1}{2}\left(\frac{v_0}{c}\right)^2 f_0\approx 2.94\times 10^{-7}f_0, \label{deltafv}
\ee
where $v_0\simeq 220$ km/s is the velocity at which dark matter orbits the center of our galaxy, i.e. the virial velocity \cite{smith2007rave}, and the frequency $f_0$ is:

\be
f_0=\frac{m_A c^2}{2\pi\hbar}. \label{f0dp} 
\ee

Dark photons cause small motions of an interferometer's mirrors, and lead to an observable effect in two ways. Firstly, the mirrors are well-separated from each other and hence experience slightly different dark photon dark matter phases. Such a phase difference leads to a differential change of the arm length, suppressed by $v_0/c$. A simple relation between dark photon parameters and the effective strain $h_D$ can be written as \cite{PierceRilesZhao2018}:

\bea
\sqrt{\left\langle h_{D}^{2}\right\rangle} &=&C\frac{q}{M}\frac{v_{0}}{2\pi c^{2}} \sqrt{\frac{2\rho_{\mathrm{DM}}}{\epsilon_{0}}}\frac{e\epsilon}{f_{0}} \nonumber  \\
&\simeq& 6.56\times 10^{-27}\left(\frac{\epsilon}{10^{-23}}\right)\left(\frac{100 \text{ Hz}}{f_0}\right),
\label{h000}
\eea
where $\epsilon_0$ is the permittivity of free space, and $C=\sqrt{2}/3$ is a geometrical factor obtained by averaging over all possible dark photon propagation and polarization directions. Equation \ref{h000} can be derived by integrating equation \ref{eqnacce} twice over time, dividing by the arm length of the interferometer, and performing the averages over time and the dark photon polarization and propagation directions.

Secondly, the common motion of the interferometer mirrors, induced by the dark photon dark matter background, can lead to an observable signal because of the finite travel time of the laser light in the interferometer arms. The light will hit the mirrors at different times during their common motions, and although the common motions do not change the instantaneous arm length, they can lead to a longer round-trip travel time for the light, equivalent to arm lengthening, and therefore an apparent differential strain~\cite{Morisaki:2020gui}. Instead of being suppressed by $v_{0}/c$ as shown in equation \ref{h000}, such an effect suffers from a suppression factor of $(f_0 L/c)$, where $L$ is the arm length of the interferometers. Similarly to equation \ref{h000}, the common motion induces an observable signal with an effective strain $h_C$ as:
\bea
\sqrt{\langle h_C^2\rangle}&=&\frac{\sqrt 3}{ 2} \sqrt{\langle h_D^2\rangle}\frac{2\pi f_0 L}{v_0}, \nonumber \\
&\simeq& 6.58\times 10^{-26}\left(\frac{\epsilon}{10^{-23}}\right).
\label{h000p}
\eea
$h_D$ maps to $h_2$ in \cite{Morisaki:2020gui}, and $h_C$ is the result of a Taylor expansion of $h_1$ in \cite{Morisaki:2020gui}. 
The interference between the two contributions to the strain averages to zero over time, which indicates that the total effective strain can be written as $\langle h_{{\rm total}}^2\rangle = \langle h_D^2\rangle+\langle h_C^2\rangle$. 

\section{Search method} \label{sec:method}

\subsection{Cross-Correlation} \label{subsec:cross-correlation}

Cross-correlation has been widely used in \gwh searches \cite{Romano:2016dpx,Abbott:2021xxi,Abbott:2021jel}, but is employed differently here. Because we are interested in ultralight dark matter, the coherence length of a dark photon signal, given by equation 2 in \cite{Miller:2020vsl}, is always much larger than the separation between earth-based detectors \cite{pierce2019dark}. 
Therefore, the interferometers should experience almost the same dark photon dark matter field, and the signals at any two detectors are highly correlated \cite{pierce2019dark}. 

Because the dark photon signal is quasi-monochromatic, we analyze the frequency domain by Discrete Fourier Transforming the strain time series. Given a total coincident observation time, $T_{\rm obs}$, for two detectors, we divide the time series into $N_{\rm FFT}$ smaller segments, with durations $T_{\rm FFT}$, i.e. $T_{\rm obs}=N_{\rm FFT}T_{\rm FFT}$. For the $i$-th time segment, $j$-th frequency bin, and interferometer $k$ (1 or 2), we label the complex Discrete Fourier Transform coefficients as $z_{k,ij}$. 
The one-sided power spectral densities (PSDs) of interferometer 1(2) can be estimated by taking a {(bias-corrected)} running median of the raw noise powers $P_{k,ij}$ from 50 neighboring frequency bins: ${\rm PSD}_{k,ij}=2{\rm P}_{k,ij}/T_{\rm FFT}$.

The cross-correlated signal strength is:
\be
S_j=\frac{1}{N_{\rm FFT}}\sum_{i=1}^{N_{\rm FFT}}\frac{z_{1,ij}z_{2,ij}^*}{P_{1,ij}P_{2,ij}},
\ee
where ``$*$'' is the complex conjugate, and the variance is:
\be
\sigma_j^2=\frac{1}{N_{\rm FFT}}\left\langle\frac{1}{2P_{1,ij}P_{2,ij}}\right\rangle_{N_{\rm FFT}},
\ee
where $\langle...\rangle_{N_{\rm FFT}}$ is the average over $N_{\rm FFT}$ time segments. Therefore, the signal-to-noise ratio (SNR) is:
\be
{\rm SNR}_j=\frac{S_j}{\sigma_j}.
\ee
{In Gaussian noise without a signal, $SNR_j$ has zero mean and unit variance.} {The presence of a signal would lead to a non-zero offset in the mean SNR proportional to $\epsilon^2$ (see equations \ref{h000}-\ref{h000p}).} We note that we will include the overlap reduction function (ORF) in our upper limit calculation, which accounts for the relative orientation and overlap of two detectors and the responses of the detectors to a signal. As indicated in \cite{PierceRilesZhao2018}, the ORF is constant ($\sim -0.9$) 
for the LIGO Hanford (H1) and LIGO Livingston (L1) detectors because the dark photon coherence length always exceeds the detector separation, {but the ORF is suppressed (-0.18) for the common-mode finite-speed effect \cite{Manita:2023mnc}, degrading the sensitivity of the cross-correlation method for this contribution.}

Here, we analyze only time segments satisfying standard data quality requirements used in \gwh searches (see section \ref{sec:data}), and {further restrict to contiguous, coincident intervals of good data spanning the Fast Fourier Transform coherence time}. As in the analysis performed using data from the first observing run (O1)~\cite{guo2019searching}, we set $T_{\rm FFT}=1800$ s, a pragmatic compromise between recovering signal power at high frequencies with shorter-than-optimal coherence times, and reducing noise contamination at low frequencies for longer-than-optimal coherence times.
{An important constraint at low frequencies is that requiring longer (contiguous) coherence times necessarily reduces total available livetime, especially given the need for coincident H1 and L1 data}. In total, we analyze 7539 pairs of 1800-second coincident time segments from H1 and L1.

\subsection{BSD analysis}\label{subsec:adaptive}

In addition to cross-correlation, we employ an independent method \cite{Miller:2020vsl} to search for dark photon dark matter. The method relies on Band Sampled Data (BSD) structures, which store the detector's downsampled strain data as a reduced analytic signal \cite{piccinibsd} in 10-Hz/1-month chunks. In each 10-Hz band, we change the Fast Fourier Transform coherence time \cite{piccinibsd} based on the expected Maxwell-Boltzmann frequency spread of dark photons, equation \ref{deltafv}.
Although this frequency spread is given as a function of $v_0$, we instead use the escape velocity from the galaxy, $v_{\rm esc}\simeq 540$ km/s \cite{smith2007rave}, to determine the maximum allowed $\TFFT$, $\tfftmax$, by requiring that the frequency spread be contained in one frequency bin in $\tfftmax$:

\be
\tfftmax\lesssim \frac{2}{f_0}\frac{c^2}{v_\text{esc}^2}\simeq \frac{6\times 10^5}{f_0} \text{ s.} \label{tfftmax}
\ee
Based on simulations \cite{Miller:2020vsl}, we found that the sensitivity of the search improves when taking $\TFFT=1.5\tfftmax$, because the power lost due to over-resolving in frequency is less than that gained by increasing $\TFFT$.

After selecting $\TFFT$, we create time/frequency ``peakmaps'' \cite{Astone:2014esa,sfdb_paper}, which are collections of ones and zeros that represent when the power in particular frequency bins has exceeded a threshold in the equalized spectrum. Because we choose $\TFFT$ to confine a signal's power to one frequency bin, we project the peakmap onto the frequency axis and look for frequency bins with large numbers of peaks, which we call the ``number count''. 

\begin{figure}
    \centering
    \includegraphics[width=\columnwidth]{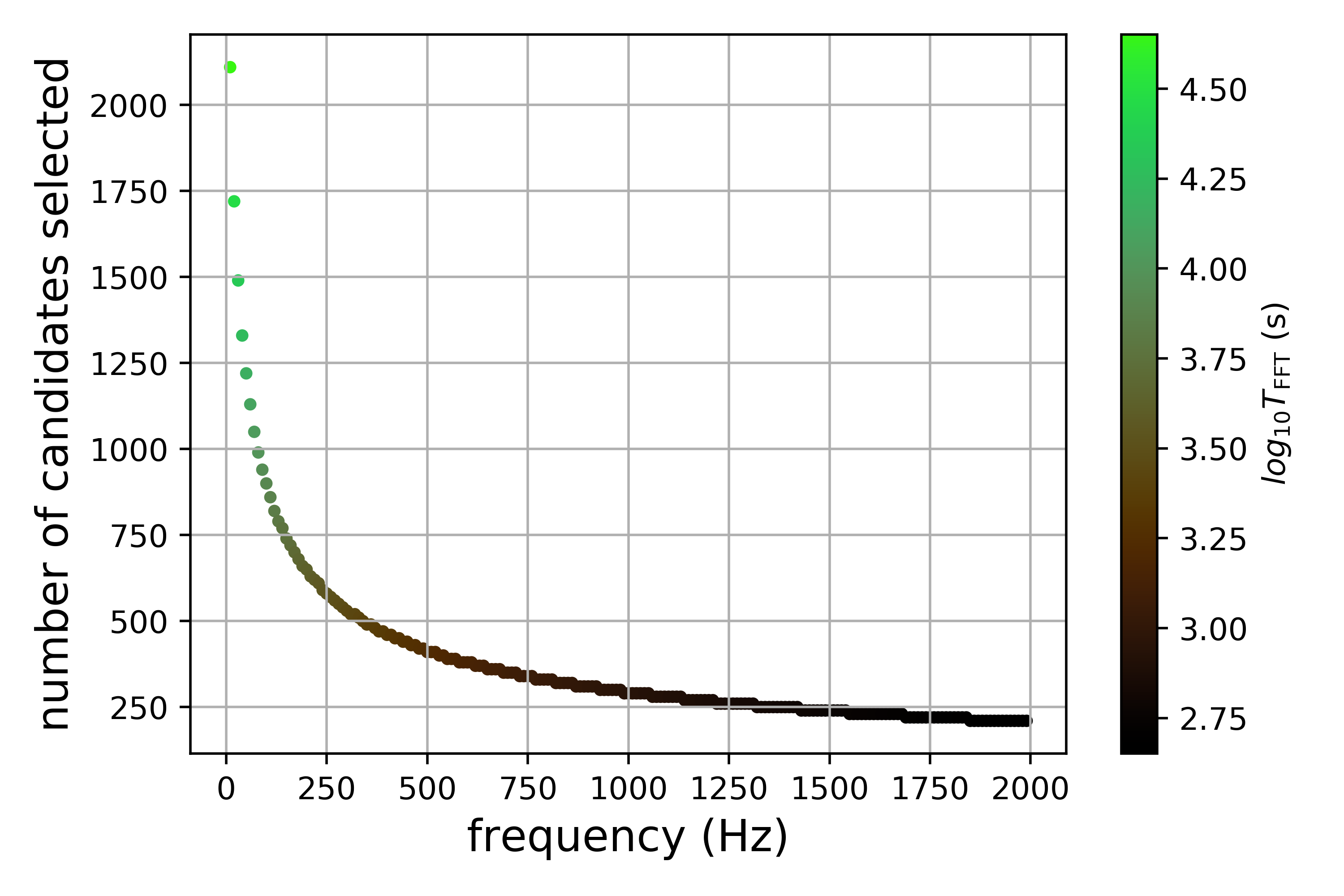}
    \caption{Number of candidates selected as a function of frequency in the BSD analysis, with $\log_{10}\TFFT$ colored. We select enough candidates in each 1-Hz band such that one coincident candidate between two detectors would occur in Gaussian noise. The changing number of candidates as a function of frequency ensures that we select uniformly in frequency. }
    \label{fig:Ncandtfft}
\end{figure}

We should uniformly select candidates in the frequency domain. 
In figure \ref{fig:Ncandtfft} shows how many candidates to select in each 10-Hz frequency band such that we would obtain, on average, one coincident candidate every one Hz in Gaussian noise. We also show in color how $\log_{10}\TFFT$ changes with frequency (equation \ref{tfftmax}).

Our detection statistic is the critical ratio $CR$:

\begin{equation}
    CR=\frac{y-\mu}{\sigma},
\end{equation}
where $y$ is the number count in a particular frequency bin, and $\mu$ and $\sigma$ are the mean and standard deviations of the number counts across all frequency bins in the band. {The $CR$ has a normal distribution with an expectation value of 0 and unit variance in Gaussian noise, and a normalized non-central $\chi^2$ distribution with two degrees of freedom when a signal is present.}
\section{Data}\label{sec:data}
We use data from the third observing run (O3)
of the Advanced LIGO \citep{2015CQGra..32g4001L} and Virgo \citep{2015CQGra..32b4001A} \gwh
detectors between 10-2000 Hz. 
O3 lasted from 2019 April 1 to 2020 March 27, with a one-month pause in data collection in October 2019. The three detectors' datasets, H1, L1, and Virgo (V1), had duty factors of $\sim 76\%$, $\sim 77\%$, and $\sim 76\%$, respectively, during O3.

In the event of a detection, calibration uncertainties would limit our ability to provide robust estimates of the coupling of dark matter to the interferometers. Even without a detection, these uncertainties affect the estimated instruments' sensitivities and inferred upper limits. The uncertainties vary over the course of a run but do not change by large values, so we do not consider time-dependent calibration uncertainties here \cite{Sun:2020wke}.

For the LIGO O3 data set, the analyses use the ``C01'' calibration, which has estimated maximum amplitude and phase uncertainties of $\sim 7\%$ and $\sim 4 \deg$, respectively \citep{Sun:2020wke}. Because of the presence of a large number of noise artifacts, \emph{gating} \cite{gatingDocument,y_stochasticGatingDocument} has also been applied to LIGO data. This procedure applies an inverse Tukey window to LIGO data at times when the root-mean-square value of the whitened strain channel in the 25-50 Hz band or 70-110 Hz band exceeds a certain threshold. The improvements from gating are significant, as seen in stochastic and continuous gravitational-wave analyses in O3 \cite{Abbott:2021xxi}.
For the Virgo O3 dataset, we use the ``V0'' calibration with estimated maximum amplitude and phase uncertainties of 5\% and 2~$\deg$, respectively.

\section{Results}\label{sec:results}

\subsection{Cross-Correlation}

The output of the cross-correlation analysis is a value of the SNR in every frequency bin analyzed. At this point, we remove frequency bins with noise artifacts, i.e. bins within 0.056 Hz of known noise lines \cite{O3lines}. To further estimate the non-Gaussian background from artifacts, control samples are constructed using frequency lags, i.e. examining the correlations among a set of offset bins. We apply ten lags of the frequency bin offsets, i.e. (-50, -40, ..., -10, +10, ..., +50). If any frequency bin in the control sample has a $|{\rm Re(SNR)}|$ or $|{\rm Im(SNR)}|$ larger than 4.0 within 0.1 Hz of the outlier, the outlier is vetoed as contaminated by spectral leakage from a nearby non-Gaussian artifact. {We choose a band of 0.1 Hz because within that band, spectral leakage causes non-physical correlated amplitudes and phases. Furthermore, ten lags allows us to compare frequency bins that are not too far from each other to construct an estimation of the noise in the chosen frequency bin}.

After removing these  instrumental artifacts, we look for frequency bins with ${\rm Re(SNR)}$ $< -5.8$, which corresponds to an overall $\sim$ 1$\%$ false alarm probability after including the trial factor in Gaussian noise, and is negative because H1 and L1 are rotated 90 deg with respect to each other. We find no outliers that pass this threshold.

Finally, as a cross-check, between
$[5.0, 5.8]$ for $|{\rm Re(SNR)}|$ or $|{\rm Im(SNR)}|$, we find four non-vetoed outliers, which are shown in table \ref{tab:corrcands}. The number of outliers is consistent with the Gaussian noise expectation of 4.1. We consider the absolute value of the real and imaginary components of the SNR because we are checking consistency with the expected number of outliers in Gaussian noise, which does not depend on the sign of the SNR . {We show the distribution of the real and imaginary parts of the SNR in appendix \ref{app:dist}}.

\subsection{BSD analysis}

Before selecting candidates, we remove any frequencies that fall within one frequency bin of known noise lines from each detector's data \cite{O3lines}. 
We subsequently require coincident candidates between two or more detectors to be within one frequency bin of each other.
At this stage, our analyses of the Hanford-Livingston (HL), Hanford-Virgo (HV), and the Livingston-Virgo (LV) baselines return 5801, 5628, and 5592 candidates, respectively.

In all baselines, we veto coincident candidates if one of the candidates' critical ratios is less than five or one of the candidates' frequencies is too close, {i.e. within 5 bins}, to the edges of the 10 Hz-band analyzed. {The latter veto is necessary because the construction of the BSDs introduces artifacts in some bands at the edges.} For the HL baseline, we remove candidates whose critical ratios differ by more than a factor of two because the sensitivity of each interferometer is comparable, so we do not expect a dark photon signal to appear with vastly different critical ratios in each detector. In the HV and LV baselines, we reject candidates whose critical ratios in V1 are higher than those in L1 or H1 because Virgo is less sensitive than LIGO \cite{Abbott:2020qfu}. {We show distributions of $CR$ in the Hanford and Livingston detectors across all frequencies in appendix \ref{app:dist}, figures \ref{fig:H_CR} and \ref{fig:L_CR}, respectively, as well as the $CR$ distribution of the number of coincident candidates in figure \ref{fig:coin_CR}. } 

\begin{figure}
    \centering
    \includegraphics[width=\columnwidth]{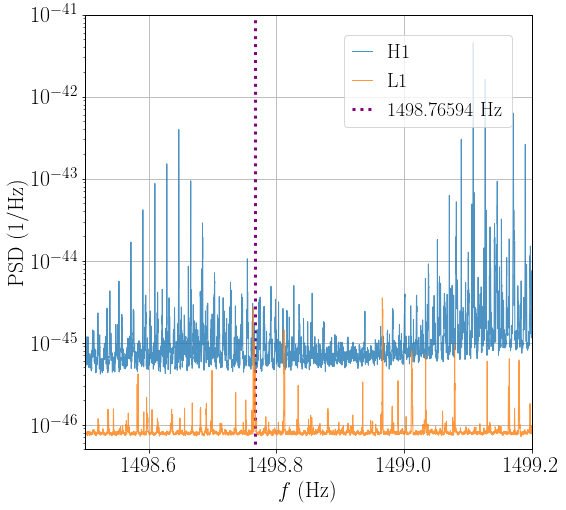}
    \caption{We discarded all surviving outliers because they were due to instrumental noise or artifacts. In this figure, we can see the comb affecting the power spectral density (PSD) of H1 and the line in L1 responsible for the production of an outlier near 1498.8 Hz. Frequency resolution: $\delta f=3.47\times 10^{-5}$ Hz. }
    \label{fig:outlier_ex}
\end{figure}

We are then left with eleven surviving candidates across the three baselines, given in table \ref{tab:cands}, that are all due to instrumental noise or artifacts in the peakmap. {Peakmap artifacts occur because when there are strong lines at particular frequencies, we tend to select peaks that correspond to those lines. This causes a ``depletion'' of peaks nearby, and thus, a candidate could result because the level of the noise in the projected peakmap is lower on one side than on the other.} No candidate has been found to be coincident in all three interferometers. These surviving candidates do not overlap with the list of known lines used in this search \cite{O3lines}, although line artifacts or/and combs regions are clearly visible when using a different resolution to construct the spectra. In figure \ref{fig:outlier_ex}, we show an example of the disturbances near an outlier at 1498.76 Hz, where a family of combs is present in both the H1 and L1 detectors.

\subsection{Upper limits}

\begin{figure*}
    \centering
    \includegraphics{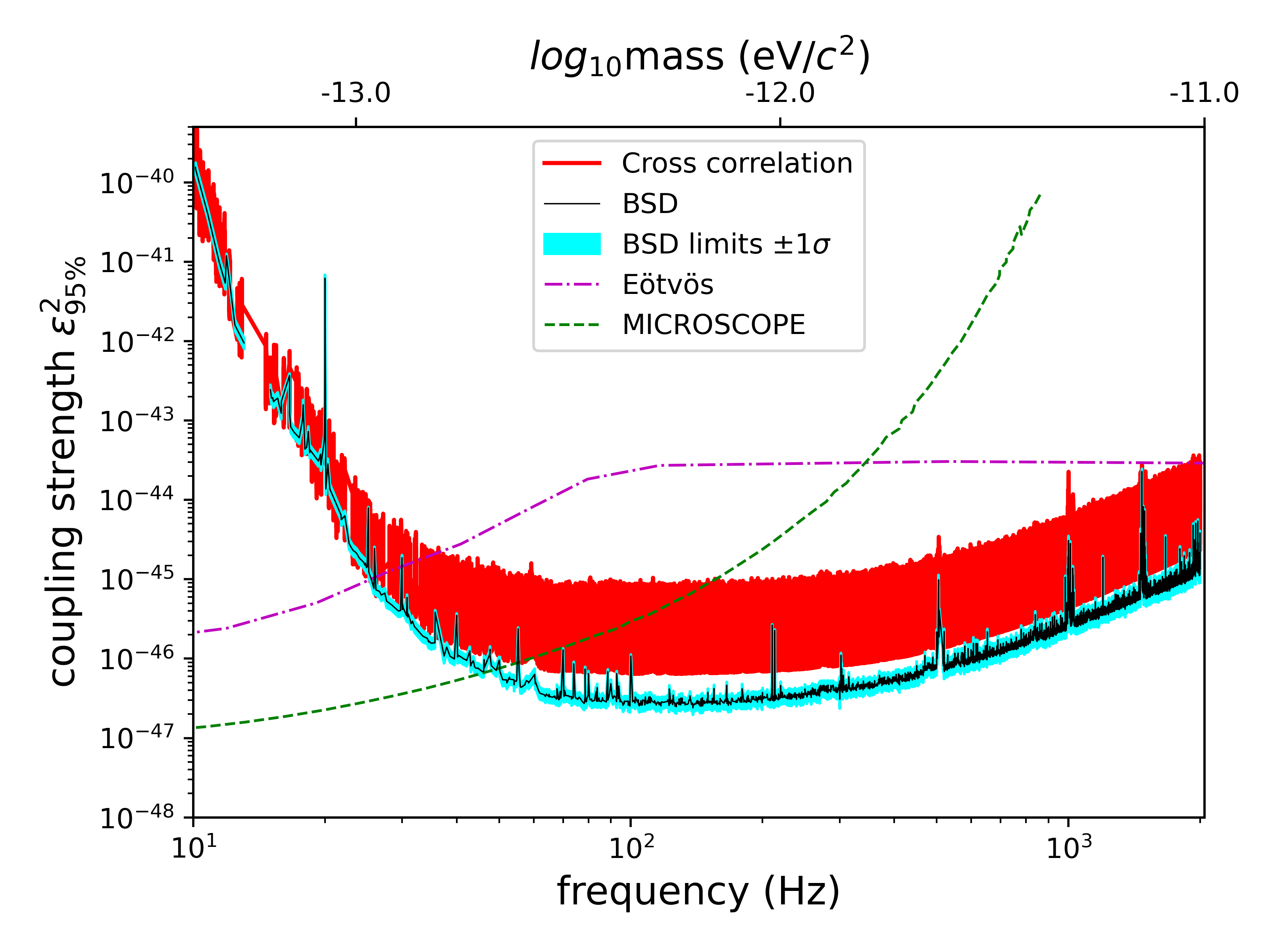}
    \caption{Upper limits derived using a Feldman-Cousins approach for both searches on dark photon/baryon coupling, $U(1)_{\rm B}$. The limits from each method are comparable, noting that the BSD-based analysis takes an optimally chosen $\TFFT$ and can observe for twice as long than the cross-correlation method can. We plot for comparison upper limits from MICROSCOPE given in
    \cite{Berge:2017ovy}, {though other weaker limits exist \cite{Touboul:2017grn,Fayet:2017pdp,Fayet:2018cjy}}, that have been converted from the coupling constant to gravity, $\alpha$, to $\epsilon^2$, using the equation below figure 3 in \cite{Michimura:2020vxn}, and from the Eötvös torsion balance experiment \cite{Schlamminger:2007ht}. To produce limits on dark photon/baryon-lepton coupling, $U(1)_{\rm B-L}$, {our} limits should be multiplied by four. {Note that this figure has been updated following an oversight in calculating the cross-correlation limits, which have been weakened by the correction discussed in  \cite{PhysRevD.109.089902}, and can be seen by comparing Fig. 3 in the previous arXiv version \cite{2105.13085v2} to this one. The BSD limits, which are unchanged, are now the most constraining at most masses. }}
    \label{fig:uls}
\end{figure*}

Finding no evidence of a signal, in figure \ref{fig:uls} we place 95\% confidence-level upper limits on the square of the minimum detectable dark photon/baryon coupling, $U(1)_{\rm B}$, using the HL baseline. The cross-correlation limits are shown in red for every 0.556-mHz bin, while the BSD limits are given in black with cyan $1\sigma$ shading {in frequency bins in which coincident candidates were found.}
To calculate these limits, we employ the Feldman-Cousins \cite{Feldman:1997qc} approach, in which we assume that both $CR$ and $SNR$ follow Gaussian distributions, 
and map the measured detection statistics to ``inferred'' positive-definite statistics based on the upper value of table 10 of \cite{Feldman:1997qc} at 95\% confidence. As shown in \cite{Miller:2020kmv}, this approach produces consistent limits with respect to those that would be obtained by injecting simulated signals. {With our estimates of the noise power spectral density and $\TFFT$, we can translate the inferred $SNR$ and $CR$ at each frequency to the corresponding signal amplitude using equation 9 in \cite{PierceRilesZhao2018} and equation 30 in \cite{Miller:2020vsl}, respectively}. This amplitude is then converted to a coupling strength using equation \ref{h000}, and adjusted for the common mode motion effect \cite{Morisaki:2020gui}. 

{The limits from the BSD method are more stringent than those from cross-correlation because the ORF used for the
contribution of the finite light speed effect is much smaller than that of the true differential motion effect, i.e. -0.18 vs. -0.9. \cite{Manita:2023mnc,PhysRevD.109.089902}. The BSD analysis does not rely on the ORF; hence, it can benefit from the full improvement in sensitivity given by the finite light travel time effect.}


\begin{table*}
\begin{tabular}{|c|c|c|c|c|c|}
\hline 
frequency (Hz) & SNR & SNR(Bkg)  \\ 
\hline 
483.872 & 0.53+5.03$i$ &  Re: [-3.62, 3.62] Im: [-3.52, 3.51]  \\ 
853.389 &  -0.18+5.02$i$ & Re: [-3.85, 3.85] Im: [-3.55, 3.90]  \\ 
1139.590 & -5.21+0.67$i$ & Re: [-3.54, 3.39] Im: [-3.61, 3.58]   \\ 
1686.598 & 5.01+1.63$i$ & Re: [-3.50, 3.70] Im: [-3.65, 3.89]    \\ 
\hline 
\end{tabular}
\caption{Four sub-threshold outliers returned by the cross correlation analysis of the HL baseline. We report the (complex) signal-to-noise ratio (SNR) for each outlier and the associated background (Bkg) SNR. For the background SNR, we include the range of the real part (Re) and imaginary part (Im) among ten lagged results. These four events are consistent with the Gaussian noise expectation over all of the clean bands in the analysis.}
\label{tab:corrcands}
\end{table*}
\begin{table*}
\begin{tabular}{|c|c|c|c|c|c|}
\hline 
frequency (Hz) & average CR & $T_{\rm FFT}$ (s) & baseline & source \\ 
\hline 
15.9000 & 5.29 &  44762 & HL & unknown line in L \\ 
17.8000 & 28.93 &  44762 & LV & unidentified line in L (17.8 Hz) \\ 
36.2000 & 8.90 & 22382 & HV & unidentified line in H (36.2 Hz) \\ 
599.324 & 12.38 & 1492 & HV & peakmap artifact; no significant candidate in L \\ 
599.325 & 12.33 & 1492 & HV & peakmap artifact; no significant candidate in L \\ 
1478.75 & 6.47 & 604 & HL & noisy spectra in H \\ 
1496.26 & 7.12 &  596 & HL & noisy violin resonance regions \\ 
1498.77 & 8.73 &  596 & HL & noisy violin resonance regions \\ 
1799.63 & 7.40 & 498 & HV & unidentified line in H (1799.63904 Hz) \\ 
1936.88 & 7.96 &  462 & HL & noisy violin resonance regions\\ 
1982.91 & 6.34 & 450 & HL & noisy violin resonance regions \\ 
\hline 
\end{tabular}
\caption{Outliers returned by the BSD analysis. The frequency resolution of each outlier is $1/T_{\rm FFT}$. We have determined the origin of all outliers to be from instrumental lines or peakmap artifacts. No outlier was found to be in triple coincidence. A list of unidentified lines can be found in \cite{O3unklines}.
}
\label{tab:cands}
\end{table*}



\section{Conclusions}\label{sec:conclusions}

We have presented strong constraints on the coupling strength of dark photon dark matter to baryons by using data from LIGO's and Virgo's third observing run. In the mass range {$m_{\rm A}\sim [2-4]\times 10^{-13}$} eV/$c^2$, we improve upon previous limits derived using data from the first observing run of LIGO \cite{guo2019searching} by a factor of {$\sim100$} in the square of the coupling strength of dark photons to baryons. This improvement is due to more sensitive detectors and to accounting for the finite light travel time \cite{Morisaki:2020gui}. Additionally, our limits surpass those of existing dark matter experiments, such as the Eötvös torsion balance and MICROSCOPE, by orders of magnitude in certain frequency bands, and support new ways to use \gwh detectors as direct probes of the existence of ultralight dark matter. As the sensitivities of current ground-based \gwh detectors improve, and third generation detectors, such as Cosmic Explorer \cite{Reitze:2019iox} and Einstein Telescope \cite{punturo2010einstein}, come online, we will dig even more deeply into the noise. Furthermore, once future-generation space-based detectors, such as DECIGO \cite{Kawamura:2020pcg}, LISA \cite{Babak:2017tow}, and TianQin \cite{Luo:2015ght}, are operational, we will probe dark photon couplings at masses as low as {$m_{\rm A}\sim 10^{-18}$} eV/$c^2$. 

\section*{Acknowledgments}
This material is based upon work supported by NSF’s LIGO Laboratory which is a major facility
fully funded by the National Science Foundation.
The authors also gratefully acknowledge the support of
the Science and Technology Facilities Council (STFC) of the
United Kingdom, the Max-Planck-Society (MPS), and the State of
Niedersachsen/Germany for support of the construction of Advanced LIGO 
and construction and operation of the GEO600 detector. 
Additional support for Advanced LIGO was provided by the Australian Research Council.
The authors gratefully acknowledge the Italian Istituto Nazionale di Fisica Nucleare (INFN),  
the French Centre National de la Recherche Scientifique (CNRS) and
the Netherlands Organization for Scientific Research, 
for the construction and operation of the Virgo detector
and the creation and support  of the EGO consortium. 
The authors also gratefully acknowledge research support from these agencies as well as by 
the Council of Scientific and Industrial Research of India, 
the Department of Science and Technology, India,
the Science \& Engineering Research Board (SERB), India,
the Ministry of Human Resource Development, India,
the Spanish Agencia Estatal de Investigaci\'on,
the Vicepresid\`encia i Conselleria d'Innovaci\'o, Recerca i Turisme and the Conselleria d'Educaci\'o i Universitat del Govern de les Illes Balears,
the Conselleria d'Innovaci\'o, Universitats, Ci\`encia i Societat Digital de la Generalitat Valenciana and
the CERCA Programme Generalitat de Catalunya, Spain,
the National Science Centre of Poland and the Foundation for Polish Science (FNP),
the Swiss National Science Foundation (SNSF),
the Russian Foundation for Basic Research, 
the Russian Science Foundation,
the European Commission,
the European Regional Development Funds (ERDF),
the Royal Society, 
the Scottish Funding Council, 
the Scottish Universities Physics Alliance, 
the Hungarian Scientific Research Fund (OTKA),
the French Lyon Institute of Origins (LIO),
the Belgian Fonds de la Recherche Scientifique (FRS-FNRS), 
Actions de Recherche Concertées (ARC) and
Fonds Wetenschappelijk Onderzoek – Vlaanderen (FWO), Belgium,
the Paris \^{I}le-de-France Region, 
the National Research, Development and Innovation Office Hungary (NKFIH), 
the National Research Foundation of Korea,
the Natural Science and Engineering Research Council Canada,
Canadian Foundation for Innovation (CFI),
the Brazilian Ministry of Science, Technology, and Innovations,
the International Center for Theoretical Physics South American Institute for Fundamental Research (ICTP-SAIFR), 
the Research Grants Council of Hong Kong,
the National Natural Science Foundation of China (NSFC),
the Leverhulme Trust, 
the Research Corporation, 
the Ministry of Science and Technology (MOST), Taiwan,
the United States Department of Energy,
and
the Kavli Foundation.
The authors gratefully acknowledge the support of the NSF, STFC, INFN and CNRS for provision of computational resources.

This work was supported by MEXT, JSPS Leading-edge Research Infrastructure Program, JSPS Grant-in-Aid for Specially Promoted Research 26000005, JSPS Grant-in-Aid for Scientific Research on Innovative Areas 2905: JP17H06358, JP17H06361 and JP17H06364, JSPS Core-to-Core Program A. Advanced Research Networks, JSPS Grant-in-Aid for Scientific Research (S) 17H06133, the joint research program of the Institute for Cosmic Ray Research, University of Tokyo, National Research Foundation (NRF) and Computing Infrastructure Project of KISTI-GSDC in Korea, Academia Sinica (AS), AS Grid Center (ASGC) and the Ministry of Science and Technology (MoST) in Taiwan under grants including AS-CDA-105-M06, Advanced Technology Center (ATC) of NAOJ, and Mechanical Engineering Center of KEK. 


{\it We would like to thank all of the essential workers who put their health at risk during the COVID-19 pandemic, without whom we would not have been able to complete this work.}

\appendix 

\section{Distribution of detection statistics}
\label{app:dist}
We provide here more details on our detection statistics for both methods. When we calculate upper limits, we assume that these statistics follow Gaussian distributions, which is actually true only in clean bands. But, because we showed the Feldman-Cousins approach to be robust towards noise disturbances in \cite{Miller:2020vsl}, we are confident that the limits are reflective of what we would have obtained if we performed software injections.

\begin{figure}
    \centering
\includegraphics[width=0.9\columnwidth]{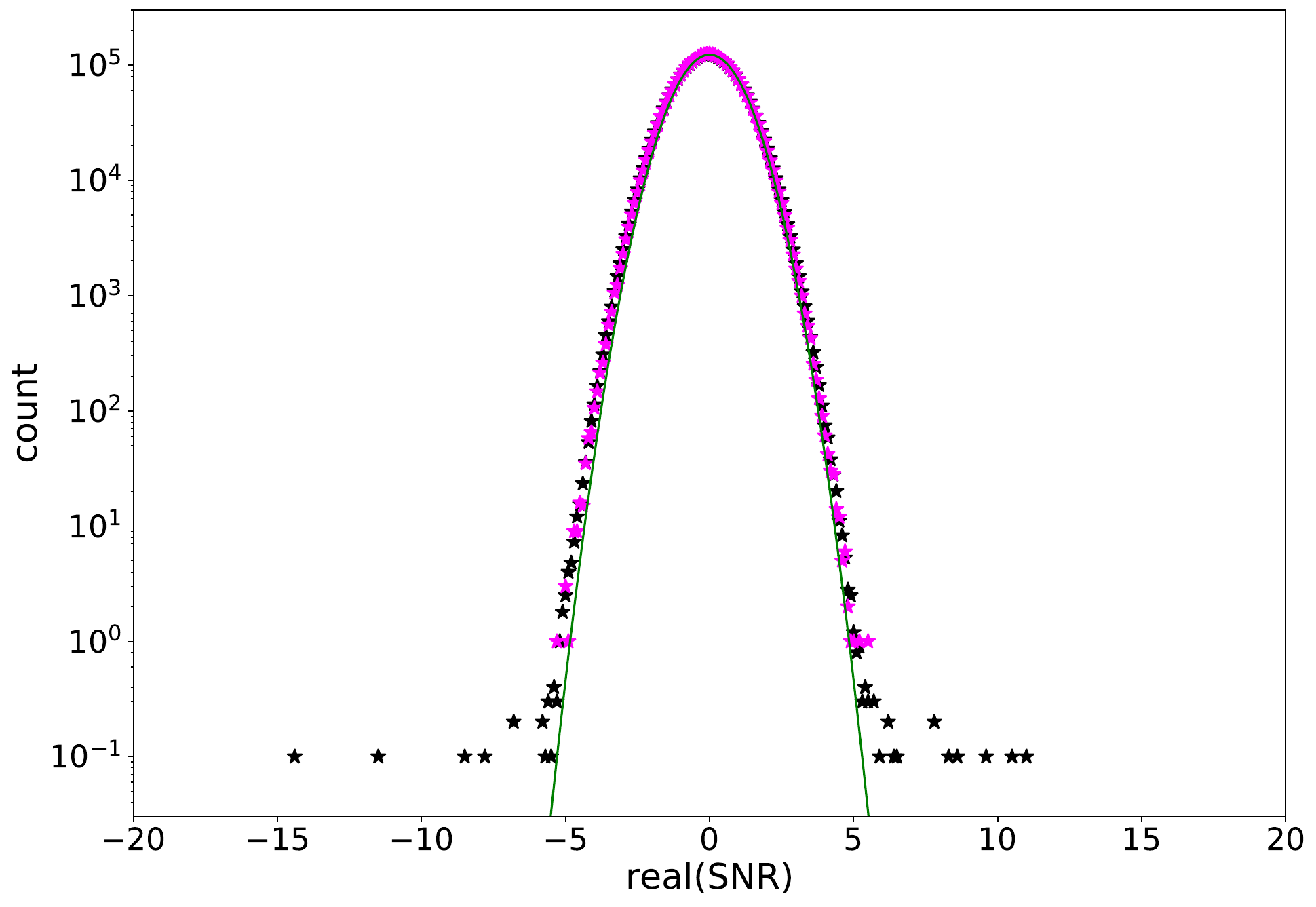}
\includegraphics[width=0.9\columnwidth]{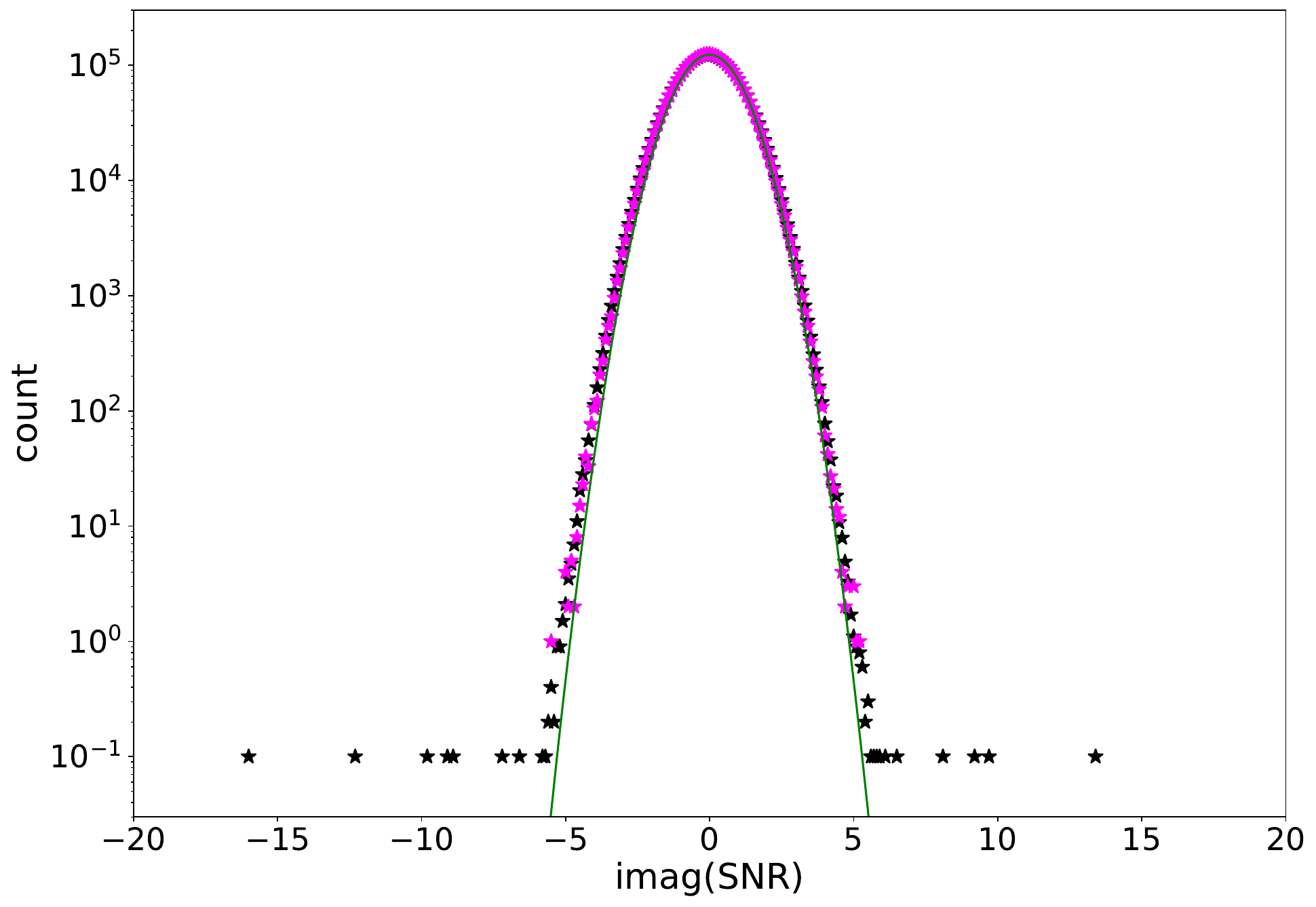}
    \caption{Distribution of the real (top) and imaginary (bottom) parts of the SNR in the cross-correlation search,
    with those corresponding to on-source (with zero lag) results in magenta, background (with frequency lags) in black and 
    the ideal Gaussian distribution in green.}
    \label{fig:snr}
\end{figure}

For the cross-correlation search, the distributions of the real and imaginary parts of the SNR are shown in figure \ref{fig:snr} after vetoing frequency bins within 0.056Hz of the known noise lines~\cite{O3lines} and after vetoing the instrumental artifacts as described in the main text above.

\begin{figure}
    \centering
\includegraphics[width=\columnwidth]{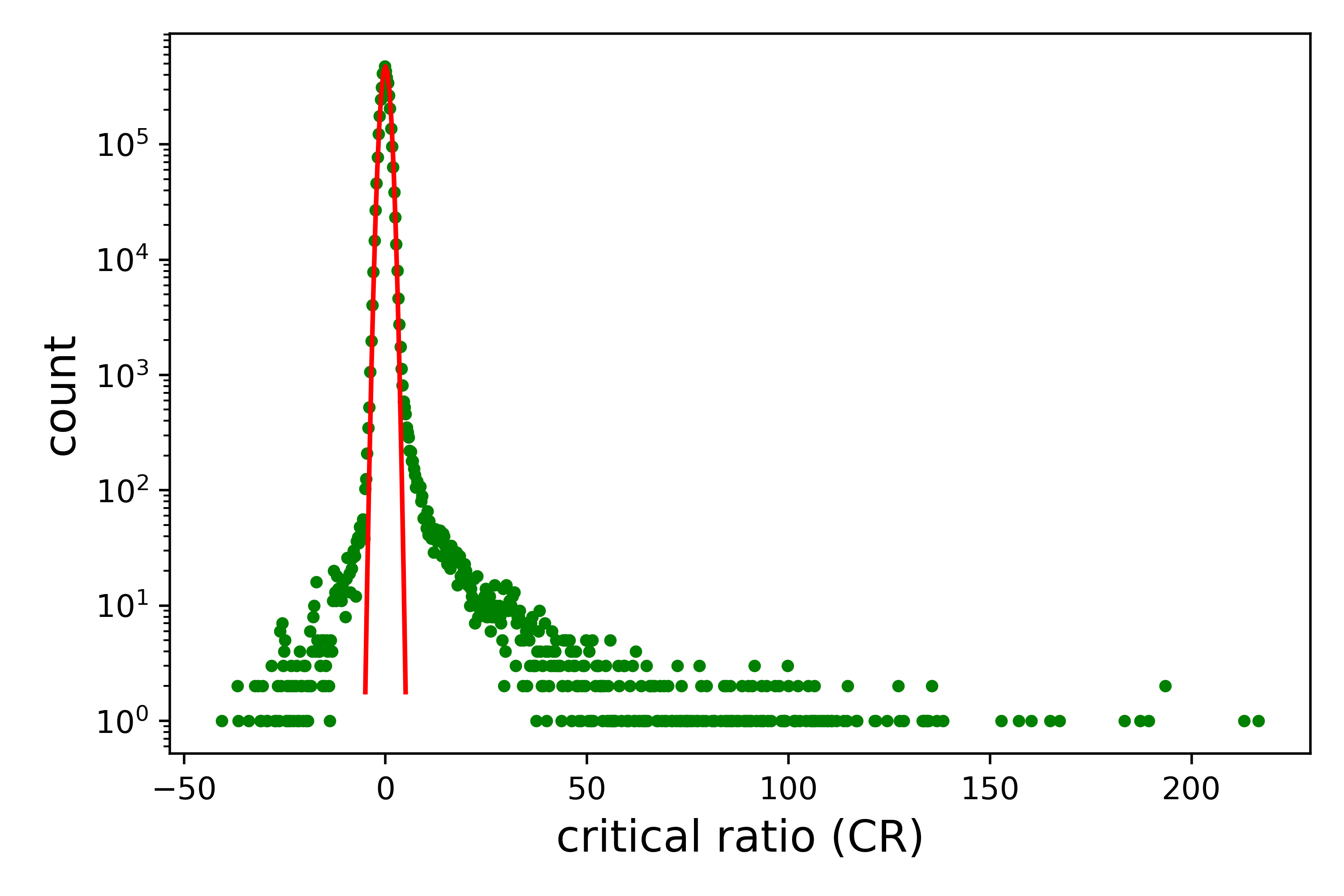}
    \caption{Histogram of critical ratios in all frequency bins in the Hanford detector, with a Gaussian (in red) overlayed. }
    \label{fig:H_CR}
\end{figure}

\begin{figure}
    \centering
\includegraphics[width=\columnwidth]{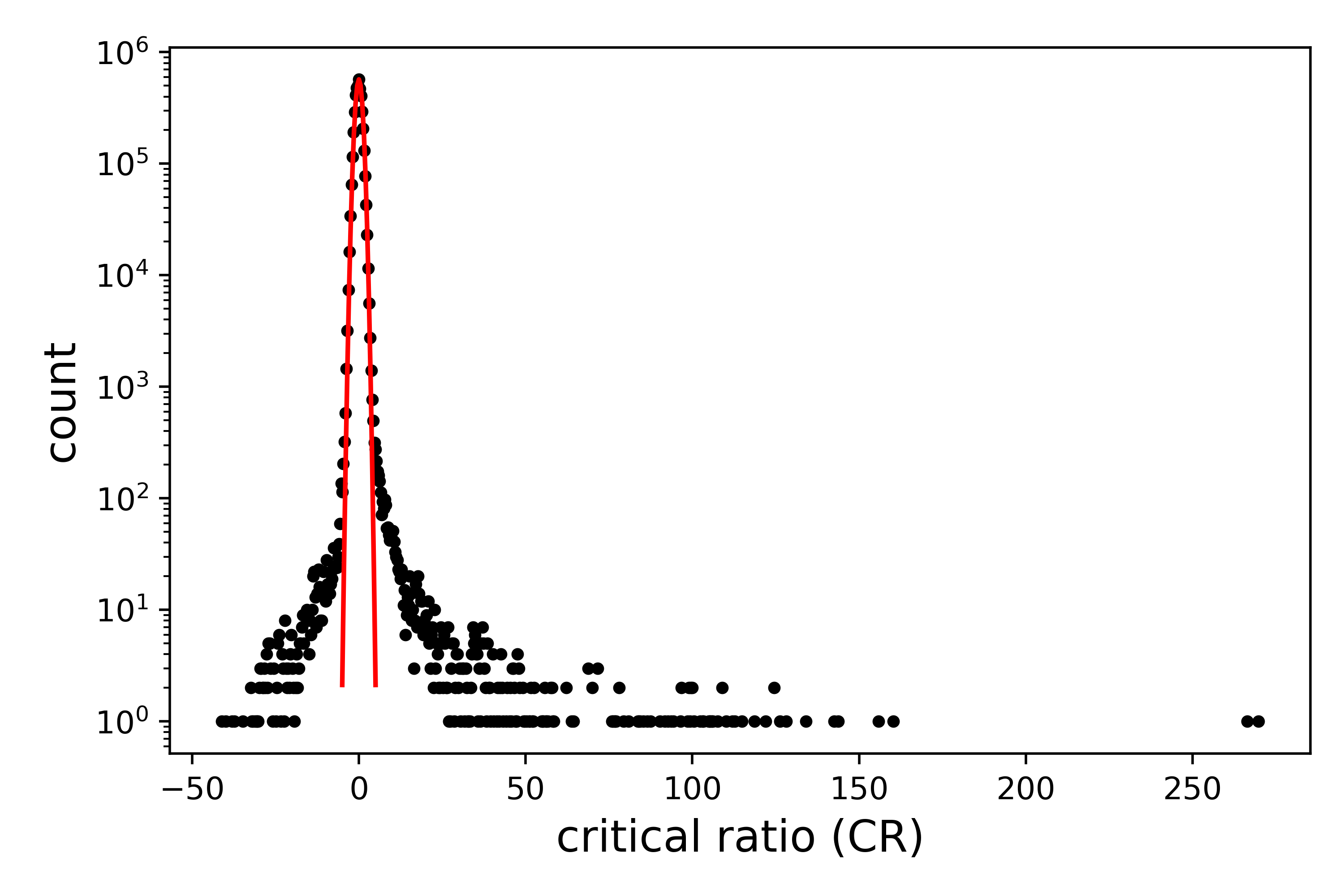}
    \caption{Histogram of critical ratios in all frequency bins in the Livingston detector, with a Gaussian (in red) overlayed. }
    \label{fig:L_CR}
\end{figure}

We show the distributions of the CR in Hanford (figure \ref{fig:H_CR}) and Livingston (figure \ref{fig:L_CR}), over all frequency bins between 10-2000 Hz. We also overlay a Gaussian on the plot to show the extent to which the distributions differs from a Gaussian distribution. In both detectors, the number of frequency bins whose CRs deviate from Gaussianity is of $\mathcal{O}(10^2)$, which is a small fraction of the total number of bins analyzed.

\begin{figure}
    \centering
\includegraphics[width=\columnwidth]{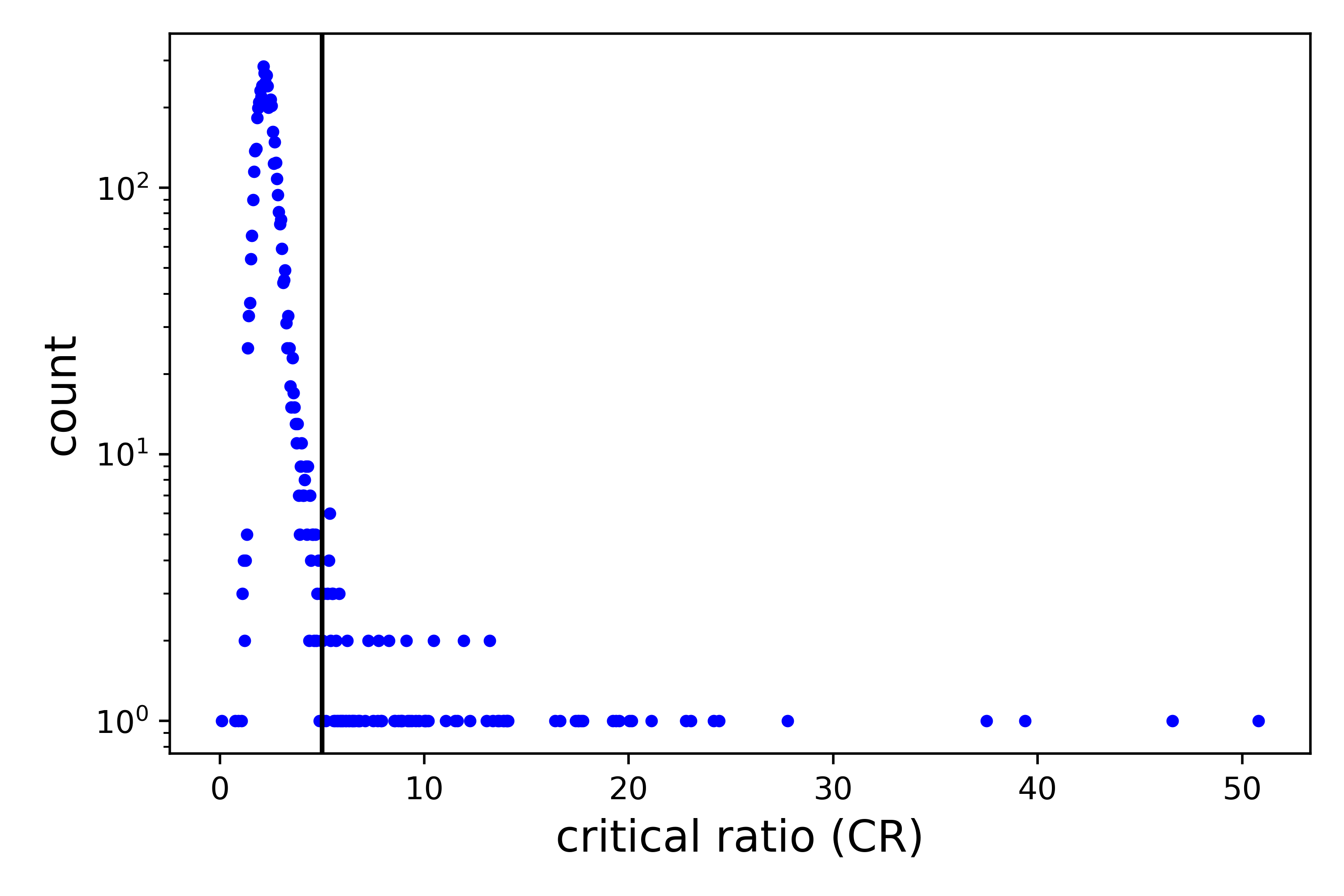}
    \caption{Histogram of coincident critical ratios, after our selection of candidates in each 10-Hz frequency band. We performed the coincidences between the candidates returned after analyzing Hanford and Livingston data. }
    \label{fig:coin_CR}
\end{figure}

We also include a plot to characterize the \emph{coincident} candidates between Hanford and Livingston that are selected in our search. Figure \ref{fig:coin_CR} shows a histogram of all the coincident candidates' critical ratios that we select, as well as a black line that indicates the threshold on the critical ratio that we impose, equal to 5. We can see that very few candidates are coincident relative to the number of candidates plotted in figures \ref{fig:H_CR} and \ref{fig:L_CR}, and that the total number of coincident candidates that are subject to further study is of $\mathcal{O}(1000)$.

\bibliographystyle{apsrev4-1}
\bibliography{references} 

\iftoggle{endauthorlist}{
  \let\author\myauthor
  \let\affiliation\myaffiliation
  \let\maketitle\mymaketitle
  \title{The LIGO Scientific Collaboration, Virgo Collaboration, and KAGRA Collaboration  }
  \pacs{}


\author{R.~Abbott}
\affiliation{LIGO Laboratory, California Institute of Technology, Pasadena, CA 91125, USA}
\author{T.~D.~Abbott}
\affiliation{Louisiana State University, Baton Rouge, LA 70803, USA}
\author{F.~Acernese}
\affiliation{Dipartimento di Farmacia, Universit\`a di Salerno, I-84084 Fisciano, Salerno, Italy}
\affiliation{INFN, Sezione di Napoli, Complesso Universitario di Monte S. Angelo, I-80126 Napoli, Italy}
\author{K.~Ackley}
\affiliation{OzGrav, School of Physics \& Astronomy, Monash University, Clayton 3800, Victoria, Australia}
\author{C.~Adams}
\affiliation{LIGO Livingston Observatory, Livingston, LA 70754, USA}
\author{N.~Adhikari}
\affiliation{University of Wisconsin-Milwaukee, Milwaukee, WI 53201, USA}
\author{R.~X.~Adhikari}
\affiliation{LIGO Laboratory, California Institute of Technology, Pasadena, CA 91125, USA}
\author{V.~B.~Adya}
\affiliation{OzGrav, Australian National University, Canberra, Australian Capital Territory 0200, Australia}
\author{C.~Affeldt}
\affiliation{Max Planck Institute for Gravitational Physics (Albert Einstein Institute), D-30167 Hannover, Germany}
\affiliation{Leibniz Universit\"at Hannover, D-30167 Hannover, Germany}
\author{D.~Agarwal}
\affiliation{Inter-University Centre for Astronomy and Astrophysics, Pune 411007, India}
% ...

\collaboration{The LIGO Scientific Collaboration, the Virgo Collaboration, and the KAGRA Collaboration}



  \newpage
  \maketitle
}

\end{document}